\documentclass[a4paper]{jpconf}
\usepackage{graphicx}
\newcommand{\za}{\alpha}
\newcommand{\zb}{\beta}
\newcommand{\ssc}{\scriptscriptstyle}
\def\bea{\begin{eqnarray}}
\def\eea{\end{eqnarray}}
\def\lla{\left\langle}
\def\rra{\right\rangle}

\begin{document}

\def\beq{\begin{equation}}
\def\eeq{\end{equation}}
\def\beqa{\begin{eqnarray}}
\def\eeqa{\end{eqnarray}}
\def\lla{\left\langle}
\def\rra{\right\rangle}
\def\za{\alpha}
\def\zb{\beta}
\def\ssc{\scriptscriptstyle}
\def\bsg{$b \to s \, \gamma\,$}
\def\ga{\mathrel{\raise.3ex\hbox{$>$\kern-.75em\lower1ex\hbox{$\sim$}}}}
\def\la{\mathrel{\raise.3ex\hbox{$<$\kern-.75em\lower1ex\hbox{$\sim$}}}}

\thispagestyle{empty}
\begin{flushright}
NCU-HEP-k069  \\
Feb 2017
\end{flushright}

\vspace*{.5in}

\begin{center}
{\bf Model of the Physical Space from Quantum Mechanics 
}\\
\vspace*{.5in}
{\textbf  Otto C.W. Kong}\\[.05in]
{\it Department of Physics and Center for Mathematics and
Theoretical Physics,\\
National Central University,  Chung-li, TAIWAN 32054 \\
E-mail: otto@phy.ncu.edu.tw}
\vspace*{1.in}
\end{center}
{\textbf Abstract :}\ \
{The physical world is quantum. However, our description of the quantum physics
still relies much on concepts in classical physics and in some cases with `quantized' 
interpretations. The most important case example is that of spacetime. We
examine the picture of the physical space as described by simple, so-called
non-relativisitic, quantum mechanics instead of assuming the Newtonian 
model. The key perspective is that of  (relativity) symmetry representation,
and the idea that the physical space is to be identified as the configuration
space for a free particle. Parallel to the case of the phase space,  we have 
a model of the quantum physical space which reduces to the Newtonian
classical model under the classical limit. The latter is to be obtained as
a contraction limit  of  the relativity symmetry.
}


\vfill
\noindent --------------- \\
$^\star$ Talk presented  at the Eighth International Workshop DICE 2016 ---
Spacetime, Matter,Quantum Mechanics (Sep 12 - 16), Castiglioncello, Italy.

\clearpage
\addtocounter{page}{-1}


\title{Model of the Physical Space from Quantum Mechanics}
\author{Otto C W  Kong}

\address{Department of Physics and Center for Mathematics and
Theoretical Physics, National Central University, Chung-li, Taiwan 32054}

\ead{otto@phy.ncu.edu.tw}

\begin{abstract}
The physical world is quantum. However, our description of the quantum physics
still relies much on concepts in classical physics and in some cases with `quantized' 
interpretations. The most important case example is that of spacetime. We
examine the picture of the physical space as described by simple, so-called
non-relativisitic, quantum mechanics instead of assuming the Newtonian 
model. The key perspective is that of  (relativity) symmetry representation,
and the idea that the physical space is to be identified as the configuration
space for a free particle. Parallel to the case of the phase space,  we have 
a model of the quantum physical space which reduces to the Newtonian
classical model under the classical limit. The latter is to be obtained as
a contraction limit  of  the relativity symmetry.
\end{abstract}

\section{Conceptual Introduction}
Newtonian mechanics is our first comprehensive model of a fundamental
particle dynamics. Most of our theoretical conceptual notions in physics
are tied to its formulation. As such, they are classical concepts. The physical 
world is quantum. Its description should be based on quantum concepts
each of which may be very different from its classical counterpart, if there 
is at all a counterpart. Most if not all of our fundamental conceptual notions
in physics as an intuitive, common sense, origin. However, such intuitive
concepts do not fix their possible mathematical formulation which is an 
abstraction necessary for the analysis in a physical theory with precision.
The formulation of such intuitive concepts into the concepts of classical
physics is what physicists are familiar with, but not at all {\em a priori}
more intuitive than their quantum counterparts. The unfortunate
common miss-statement of quantum physics being less than intuitive 
has however hindered our appreciation of quantum physics with the
more appropriate quantum formulation of the fundamental conceptual 
notions. Our model of space and time is the most central problem at
hand. A common theme in the classical concepts is that the corresponding
physical quantities, such as the position of a particle in space, have real 
number values.  Quantum physics indicates otherwise, at least not 
necessarily by a finite number of real numbers. More generally, one
should consider such quantities as having values modeled by some
appropriate types of abstract algebraic entities. The system of real 
numbers as one of the latter has no preferred role apart from being our
first order approximation to the physical nature as in the classical theory.
We can summarize it all by saying that the main culprit in the limitation
of the classical concept is the real number, which the philosopher 
W.~Quine once called a `convenient fiction'; with quantum physics, 
the old fiction is not so convenient any more.

Let us focus on the notion of the physical space as a collection of all
possible positions, either of a free particle or equivalently as the center
of mass for a system of particles. The first mathematical model is given
by the three dimensional Euclidean geometry, which had been the only
available model for centuries till beyond Newton's time. And it was of
course adopted by Newton in his mechanics. Again, any
mathematical model is beyond naive intuition. When, as it stand today,
we have more than one such model, which one is the correct, or simply
better, model for the physical space is a physics question. The answer
is to be seek from matching the corresponding theories to experimental
result. The limited success of Newtonian mechanics says that the
Newtonian model may be not good enough. We are now familiar with
the Einstein models, for his theories of special and general relativity.
However, the Einstein models are still limited to real number geometry.
Like the Newtonian model, they are classical (physical) models. We
want to look at quantum models. A simple natural guess based on 
modern mathematics would be a kind of topological space which likely 
has some noncommutative geometry \cite{ncg}. The idea of quantum
geometry is not new in the subject matter related to deep microscopic
structure of spacetime and quantum gravity. However, we present
here a model of the physical space as behind simple quantum mechanics
and illustrate its having the correct classical limit \cite{1,2}.

At this point, some readers may feel uncomfortable about our suggestion
of going beyond the notion of a three dimensional manifold calling it 
nothing counter-intuitive. The following comment is in order. The notion 
of a three `dimensional' space, or four `dimensional' spacetime, is no 
doubt quite intuitive. The Chinese term for the universe as a totality of 
all spacetime, for example, has two characters the first of which is the
space part which literally means extending along the three principle axes. 
However, it is more tri-axial than three dimensional. Nothing in our 
intuition points towards identifying a definite position along any axis, 
or equivalently a distance with the abstract mathematical entity called 
a real number. One can see that the notion of being sort of tri-axial 
stays in our model.

\section{The Role of Relativity Symmetries}
The `particle' is the key concept for an ideal physical object in the
Newtonian theory, one that still play a key role in simple quantum
mechanics as a quantized version of the latter. To analyze the logic
behind the Newtonian formulation, the first question is `why particle?'.  
The particle is matter (with a single characteristic called mass) which
has an unambiguous position in the Euclidean space. The Newton's 
laws are really definitions of notions like states (of motion), valid frames 
of reference, force, and mass (as inertia). Note that time is also 
modeled on an (one dimensional) Euclidean geometry. All the notions 
put together gives a model of physical phenomena the success of 
which obviously breaks down at the atomic scale and beyond. We 
argue that the modified theory of quantum mechanics has all the 
notions modified including those of position in space, though time 
is not touched so long as the `non-relativistic' theory is concerned.
We stick to the setting here. 

In Ref.\cite{1}, we presented the perspective of taking the relativity 
symmetry as the starting point to look at the Newtonian model as 
well as formulating the quantum model and its classical limit. It is 
a perspective that is behind a big program aiming at eventually
constructing the quantum spacetime model and its dynamics at the
deep microscopic level \cite{3,4}. Symmetry is the single most 
important theoretical structure in modern physics. A relativity symmetry 
is the key symmetry behind a theory of mechanics. It is the group of 
reference frame transformation, and the symmetry of the spacetime 
model in the theory. The configuration space and phase space of a free 
particle are all natural homogeneous spaces as representations of the 
symmetry. The algebra of observables also carries a representation 
of it. 

For the Newtonian theory, the relativity symmetry is given by the
Galilei group. Newtonian space-time, as well as the single particle
configuration and phase spaces can all be identified as coset spaces 
of the Lie group. The coset for the space-time is given by the quotient
of the Galilei group $G(3)$ factored by the $ISO(3)$ subgroup of 
rotations and boosts. The corresponding case for the Einstein theory 
is the Minkowski spacetime $M^4= ISO(1,3)/SO(1,3)$, the coset space
of the Poincar\'e group factored by the Lorentz group. In fact,
the Newtonian limit is exactly to be recovered by a contraction
of $ISO(1,3)$ to $G(3)$ as $c \to \infty$ \cite{c,067}. What we 
have, as shown below, is much of a parallel picture for the quantum
versus classical case. The action of $G(3)$ on the coset is given by
the familiar expressions
\bea
\left(\begin{array}{c}
t' \\{ x'^i} \\ 1
\end{array}\right)
=\left(\begin{array}{ccc}
1 & 0 & B \\
V^i & R^i_{\,j}  & A^i \\
0 & 0 & 1
\end{array}\right)
\left(\begin{array}{c}
t \\ x^j \\ 1
\end{array}\right)
= \left(\begin{array}{c}
  t+B \\
  V^i t +{ R^i_{\,j} x^j + A^i} \\ 1
\end{array}\right) \;,
\eea
and the more convenient infinitesimal form 
\bea
\left(\begin{array}{c}
dt \\ {dx^i} \\ 0
\end{array}\right)
=\left(\begin{array}{ccc}
0 & 0 & b \\
v^i & \omega^i_{\,j}  & a^i \\
0 & 0 & 0
\end{array}\right)
\left(\begin{array}{c}
t \\ x^j \\ 1
\end{array}\right)
= \left(\begin{array}{c}
  b \\
  v^i t + {\omega^i_{\,j} x^j + a^i} \\ 0
\end{array}\right) \;.
\eea
Kinematical structure alone is enough to look at the configuration 
and phase spaces on which generic dynamics can be described with 
a time translation generated by a Hamiltonian with a nontrivial 
potential. Hence, we first restrict our analysis the subgroup,
$G(3)$ with the time translations $T$ taken out;  denoted by $G(3)_s$.
The Newtonian space can be identified with $G(3)_s/ISO(3)$
which is the same as $G(3)/ISO(3)\! \times\! T$.
For convenience in matching with the familiar quantum picture, we 
use $X_i $ generators in place of the boost generators $K_i=m X_i$.  
The Newtonian space then has $dx^i = \omega^i_{\,j} x^j + \bar{x}^i$, 
and for the phase space $(p^i,x^i)$, we have also 
$dp^i = \omega^i_{\,j} p^j + \bar{p}^i$. The phase space is given 
by $G(3)_s/SO(3) =G(3)/SO(3)\!\times\!  T$. The position and 
momentum coordinates $x^i$  and $p^i$ both transform as a 
three vector under rotations and have their own independent 
translations generated by $P_i$  and $X_i$, respectively.

\section{The Quantum Phase Space Is The Physical Space}
The quantum symmetry differs from the classical one. We have
the Heisenberg commutation relation which corresponds to having
the Lie algebra to the Galilean symmetry modified by
\bea
[X_i, P_j] = i\delta _{ij} I \;,
\eea
where $I$ is an extra central generator that commutes with all
others. It is a bigger group $\tilde{G}(3)$ which is a $U(1)$ central
extension of $G(3)$ \cite{gq}. Without the time translations, it is a 
Heisenberg-Weyl group supplemented with the rotations 
denoted by $H_{\!\ssc R}(3)$. In the setting, one has naively the
space coset  $H_{\!\ssc R}(3)/ISO(3)$ with
\bea \label{qsc}
\left(\begin{array}{c}
dx^i  \\  d\theta \\    0
\end{array}\right) =
\left(\begin{array}{ccc}
 \omega^i_j &   0  &  \bar{x}^i\\
\bar{p}_j  & 0   &   \bar{\theta} \\   
 {0}   & {0} &  0 
\end{array}\right)
\left(\begin{array}{c}
x^j \\ \theta \\    1
\end{array}\right)
=\left(\begin{array}{c}
\omega^i_j  x^j + \bar{x}^i\\
  \bar{p}_j  x^j  +\bar{\theta} \\   0
\end{array}\right) \;,
\eea
and the phase space coset $H_{\!\ssc R}(3)/SO(3)$ with
\bea \label{qpsc}
\left(\begin{array}{c}
dp^i  \\ dx^i\\ d\theta \\    0
\end{array}\right) =
\left(\begin{array}{cccc}
 \omega^i_j &  0& 0  &  \bar{p}^i\\
0  & \omega^i_j &  0 &  \bar{x}^i \\
-\frac{1}{2}\bar{x}_j  & \frac{1}{2}\bar{p}_j  &  0 & \bar{\theta} \\
 {0}   & {0} &  0 & 0
\end{array}\right)
\left(\begin{array}{c}
p^j \\ x^j  \\ \theta \\    1
\end{array}\right)
=\left(\begin{array}{c}
\omega^i_j  p^j + \bar{p}^i\\
\omega^i_j  x^j + \bar{x}^i \\ 
  \frac{1}{2}( \bar{p}_j   x^j -\bar{x}_j p^j)  + \bar{\theta} \\   0
\end{array}\right) \;.
\eea
Note that $H_{\!\ssc R}(3)$ is the central extension of $G(3)_s$ 
above. The funny cosets with a $\theta$ coordinate apparently do 
not serve our purpose. However, we know how to construct the
quantum phase space from the second coset. That is the formulation 
of the Hilbert space of canonical coherent states \cite{cs}. 

The phase space coset is isomorphic to the Heisenberg-Weyl
subgroup $H(3)$.  A unitary representation of the latter is given
by taking the linear span of the canonical coherent states each of
which can be identified as a point in the coset. They are given by
\bea
e^{i\theta}\left|p^i,x^i\rra =  U(p^i,x^i,\theta) \left|0\rra
\eea
where
\bea
U(p^i,x^i,\theta) \equiv 
e^{i\frac{x_i p^i}{2}} 
e^{i\theta\hat{I}} 
e^{-ix^i\hat{P}_i} 
e^{ip^i\hat{X}_i} 
= e^{i(p^i\hat{X}_i- x^i\hat{P}_i +\theta\hat{I})} \;,
\eea
and $\left|0\rra \equiv \left|0,0\rra$ is a fiducial normalized vector, 
$\hat{X}_i$ and $\hat{P}_i$ are representations of the generators $X_i$ 
and $P_i$ as Hermitian operators, and $\hat{I}$ is the identity 
operator representing the central generator $I$.  As a basis, the
set of coherent states is overcomplete. Wavefunction $\phi(p^i,x^i)$
for each of the state is a symmetry minimal uncertainty Gaussian
centered at the state label, which correspond to expectation values
rather than eigenvalues of the $\hat{P}_i$ and $\hat{X}_i$ observables.
The operators generates translations of $x^i$ and $p_i$ exactly as
described by the coset representation, with the nontrivial phase
transformation. The $\left|0,0\rra$ is the same as the ground
state for a harmonic oscillator. The coherent states are like the 
classical state described in quantum mechanics.

What is particular interesting to note is that one can have the
parallel construction starting from the (configuration) space coset 
of Eq.(\ref{qsc}).The relevant subgroup is the one generated by
$P_i$ and $I$. We have a unitary representation as the span 
\bea
e^{i\theta}\left|x^i\rra = U'(x^i,\theta) \left|0\rra
\eea
where
\bea
U'(x^i,\theta) \equiv e^{i\theta\hat{I}}
e^{-ix^i\hat{P}_i}  \;,
\eea
$\left|0\rra$ is the fiducial normalized vector, and $\hat{P}_i$ 
and $\hat{I}$ the Hermitian operators generating translations 
in $x^i$ and the phase rotation. Checking from the coset space 
action, one can see that the $\left|x^i\rra$ states are position
eigenstates. We have hence a Hilbert space as the one usually
presented in basic quantum mechanics textbook, which is
of course unitary equivalent to the previous one from the
coherent states. It looks very much like it is the quantum space
we have, and the quantum configuration space for a particle
is the same as its phase space. 

The real quantum phase space is not the Hilbert space but its
projective counterpart. Each one dimensional subspace as a ray
corresponds to a distinct physical state. The projective Hilbert
space is known to have the structure of an infinite dimensional
symplectic manifold and Kahl\"er manifold \cite{B}. We can look it 
that through an expansion of each state in terms of the Fock state
basis as
\bea
\left| \phi \rra =\sum (q_n +ip_n)\left| n \rra
\eea
for real homogeneous coordinates $q_n$ and $p_n$ of the 
projective space. The Schr\"odinger equation 
\bea
i\hbar \frac{d}{dt} \left| \phi \rra = \hat{H} \left| \phi \rra
\eea
for the state is actually equivalent to the set of Hamilton
equations 
\bea &&
\frac{d}{dt}  q_n = \frac{\partial}{\partial{p_n}} H(p_n,q_n) \;,
\nonumber \\ &&
\frac{d}{dt}  p_n = -\frac{\partial}{\partial{q_n}} H(p_n,q_n) \;,
\eea
with the Hamiltonian function 
$H(p_n,q_n)=\frac{2}{\hbar} \lla \phi | \phi \rra$. The basic fact 
is unfortunately much less appreciated than it should be. The 
projective Hilbert space also has a metric.  It has no problem 
serving as a model of the physical space either.  The countable 
infinite set of  complex coordinates $q_n +ip_n$ are just 
$\lla  n|\phi\rra$. We can also think of 
$\phi(x^i)= \lla  x^i|\phi\rra$ as a set of complex coordinates.
A functional analog of the Hamilton equations can be obtained
for the real and imaginary parts  of the function. The real and 
imaginary parts  of each   $\lla  x^i|\phi\rra$ then serves as a set 
of real coordinates. It is obviously essentially the same for the 
coherent state wavefunction $\phi(p^i,x^i)=\lla  p^i,x^i|\phi\rra$.
There are uncountable infinite number of them, hence a set 
with much redundancy. The picture says a quantum wavefunction 
is a full and definite description of the position (and momentum) 
of the particle. The space of all such position a free particle  can
possibly have, {\em i.e.} the full (projective Hilbert space is a model 
of the quantum space. Our familiar position operator $\hat {X}_i$
describe a notion of position which is obviously the one in
the classical model. The quantum particle does not have a
unique position in the classical model of space for the main
reason that the latter fails as a good model of the space.  
We can also see that it is the phase transformations generated
by the central charge that makes it no longer possible to have 
a reduced notion of the (configuration) space from the phase
space. The Lagrangian submanifold of a symplectic manifold 
of a classical system, as the configuration space, can be taken 
as the real part of the symplectic manifold with a specific
complex structure.  A phase transformation that mixes the
real and imaginary parts is a canonical transformation which
mixes the configuration and momentum variable and not a
symmetry transformation of the (configuration) space itself. With
the quantum symmetry, we have phase transformations within
the relativity symmetry. They are hence symmetry transformation 
of (configuration) space. The feature is essentially dictated by the 
Heisenberg commutation relation.

\section{Classical Limit as an Approximation}
We have our quantum model of the physical space. A nontrivial 
check that the idea is admissible is that it can explain the success 
of the corresponding classical picture in the proper limit. The latter 
is given by a symmetry contraction. More specifically, one should
take the original representation which describes the quantum physics
to the required limit rather than directly building the classical physics
description from the contracted symmetry. We will see that the 
contraction of the representation does indeed give a representation 
of the contracted symmetry \cite{3,4}. Classical mechanics should 
be seen as such an approximation to quantum mechanics \cite{070}. 
We require the quantum space model to reduce to the classical one 
as sort of the $\hbar \to 0$ limit through tracing the impact of the
contraction on the Hilbert space representation(s) described. 
We are talking about the infinite dimensional curved space as
the projective Hilbert space approximated by a finite dimensional  
Euclidean space. It works !

The symmetry contraction is given on the Lie algebra as the
 $k \to \infty$ with $X_i^c=\frac{1}{k} X_i$ and  $P_i^c=\frac{1}{k} P_i$.
Naively, one can take $\frac{1}{k^2}$ as $\hbar $.  The nonzero
commutator within $H(3)$, now in terms of $X_i^c$, $P_i^c$ 
and $I$, goes as
\bea
[X_i^c,P_j^c]=\frac{i}{k^2}\delta_{ij} I  \rightarrow  0 \;.
\eea
Hence, symmetry (sub)group, as well as the full $\tilde{G}(3)$
reduces to a trivial central extension of the classical symmetry.
The $I$ generator completely decoupled from the rest and 
becomes quite irrelevant. One can see that the transformation
properties of the coset space representations in Eqs.(\ref{qsc}) 
and (\ref{qpsc}) reduce to the classical analog as 
\[
H_{\!\ssc R}(3)/SO(3) \to  G_s(3)/SO(3) \times U(1) 
\]  and \[
  H_{\!\ssc R}(3)/ISO(3)  \to  G_s(3)/ISO(3) \times U(1) \;.
\]
Explicitly, we have at the  $k \to \infty$ limit
\bea
&& d\theta=\bar{\theta} \;,     \nonumber \\
&& dx^i_c =\omega^i_j  x^j_c + \bar{x}^i_c \;,    \nonumber \\
 && dp^i_c =\omega^i_j  p^j_c + \bar{p}^i_c \;.
\eea
For the coherent state picture, we first have the $\left|p^i,x^i\rra$
states relabeled as $\left|\tilde{p}_i^c,\tilde{x}_i^c\rra$ where
$\tilde{p}_i^c = \sqrt{\hbar} p_i$ and $\tilde{x}_i^c = \sqrt{\hbar} x_i$ 
($p_i=p^i$, $x_i=x^i$). The new label correspond to the expectation
values of $\hat{X}_i^c$ and $\hat{P}_i^c$. We have then
\bea&&
\lla \tilde{p}'^c_{i},\tilde{x}'^c_{i} \right| \hat{X}_i^c \left|\tilde{p}_i^c,\tilde{x}_i^c\rra
= \frac{(\tilde{x}'^c_{i}+\tilde{x}^c_{i})-i(\tilde{p}'^c_{i}-\tilde{p}^c_{i})}{2}
{\lla \tilde{p}'^c_{i},\tilde{x}'^c_{i} |\tilde{p}_i^c,\tilde{x}_i^c\rra } \;,
\nonumber \\ &&
\lla \tilde{p}'^c_{i},\tilde{x}'^c_{i}\right| \hat{P}_i^c \left|\tilde{p}_i^c,\tilde{x}_i^c\rra
= \frac{(\tilde{p}'^c_{i}+\tilde{p}^c_{i})+i(\tilde{x}'^c_{i}-\tilde{x}^c_{i})}{2}
{\lla \tilde{p}'^c_{i},\tilde{x}'^c_{i} |\tilde{p}_i^c,\tilde{x}_i^c\rra} \;,
\eea
with
\bea &&
{\lla \tilde{p}'^c_{i},\tilde{x}'^c_{i} |\tilde{p}_i^c,\tilde{x}_i^c\rra} =
\exp\!\! \left[i\frac{\tilde{x}'^c_{i} \tilde{p}^c_{i} - \tilde{p}'^c_{i} \tilde{x}^c_{i}}{2\hbar} \right] 
{\exp\!\! \left[-\frac{(\tilde{x}'^c-\tilde{x}^c)^2+(\tilde{p}'^c-\tilde{p}^c)^2}{4\hbar}\right]}\;,
\label{e}\\ &&
 {\lla \tilde{p}^c_{i},\tilde{x}^c_{i} |\tilde{p}_i^c,\tilde{x}_i^c\rra} =   1 \;.
\eea
The contraction result correspond to the $k \to \infty$ limit, hence
$\hbar \to 0$. One can see that the last exponential factor hence 
the full result in Eq.(\ref{e}) goes to 0. The basis states becomes 
mutually orthogonal. The equation above then say that the 
$\hat{X}_i^c$ and $\hat{P}_i^c$ operators become diagonal on
the basis set. The conclusion there is that the original Hilbert space
representation becomes a reducible representation. It reduces to
a simple sun of the one dimensional spaces for the 
 $\left|\tilde{p}_i^c,\tilde{x}_i^c\rra$ states. But such rays of
the original Hilbert space each correspond to a single state. 
They are the only states remain in the classical limit. All the
nontrivial linear superpositions are eliminated. That is to say,
the original projective Hilbert space reduces to the classical
six dimensional phase space.

What about the physical space or the configuration space?
We can actually do better than simply saying the latter is 
naturally a Lagrangian subspace of the (classical) phase space.
We can start from looking at the $\left|x^i\rra$ basis states
we have in constructing the quantum Hilbert space as the
model of the physical space and trace its contraction limit.
The basis states are to be relabeled as $\left|x^c_i\rra$,
{\em i.e.} by eigenvalues of the $\hat{X}_i^c$. Now, 
$\hat{P}_i^c$ and hence all operators, say, as polynomials
of $\hat{X}_i^c$  and $\hat{P}_i^c$  commute. We have the
same conclusion for the reduction of the Hilbert space now
seeing as going to the simple sum of the rays of  $\left|x^c_i\rra$.
The projective Hilbert space is of course the three dimensional
Euclidean space.

\section{Concluding Remarks}
In physics, we are supposed to learn from experiments what constitutes 
a good/correct theoretical/mathematical model of any physical concept, 
and physical space should not be an exception. As physics advances,
we see our understanding of all the intuitive concepts keeps deepening.
The abstract mathematical contents to be given to such concepts
changes as we replace a fundamental theory with a better one. 
Spacetime, or space and time, is the most basic concepts in our all
discourse on physics. Einstein's relativity theories advanced our 
understanding of spacetime. Looking at it from the proper point of
view, quantum mechanics does the same. In a way, it would be quite
surprising if it does not. Our talk here offers exactly such a perspective. 

A notion about spacetime being somehow emergent rather than
fundamental has been getting popular lately. In many cases, it is not
clear what the author(s) exactly meant about that. The way we see it:
The familiar classical picture and all the related features are almost
definitely not at all fundamental. They are like our classical 
approximate description of spacetime. It should be quite clear
nowadays that the deep microscopic picture of spacetime has to 
be very different. We sure expect any good model of quantum
spacetime model to give back the classical model as an 
approximation under proper limits. As such, one can certainly
call the classical model a emergent one. However, to have a
fundamental theory of physics built completely without some
notion of spacetime in it, we think, would be quite impossible.
There has to be some dynamical variables, whatever unfamiliar
mathematical objects needed to described them, in the theory.
For the part of such variables necessary to give a picture of
the spacetime as we have in the known classical theories at the
proper limit, it is only fair to call the collection of all their
admissible values in the most general physical setting the
spacetime model of the theory. 

We would also like to point out that quantum field theory 
understood correctly gives another very interesting 
perspective about spacetime, a feature we expect to stay
as a part of any model of the deep microscopic (quantum)
spacetime. Quantum field theory says that spacetime is the
only physical entity. The various quantum fields are not
independent dynamical `objects' but only dynamic degrees
of freedom the spacetime has. Any state, says the state 
which corresponds to a single muon, involves all the
degrees of freedom. With enough precision,  we can see
the role of any degree of freedom in any other basic 
properties of the state. The magnetic dipole moment of
the muon has been used to reveal information about 
possible superparticles --- postulated new quantum fields
the discovery of which is one of the key target of the 
Large Hadron Collider experiments. We know already from
quantum entanglement that the full system is more than
the sum of its parts. There are no truly independent
subsystems.  

The above leads naturally to the question of gravitation
or quantum gravity. We believe {\em `Noncommutative 
Geometry is to Quantum Gravity as NonEuclidean 
Geometry is to Gravity'}. One should build quantum gravity 
as a {\em geometrodynamics of quantum spacetime}.
Our quantum space model presented here is of course
very far from the final quantum spacetime model. We
need to first go from the $\tilde{G}(3)$ setting to one that
incorporates Poincar\'e symmetry.  Actually, we have a basic 
perspective which goes beyond that to a full stable symmetry
that cannot be the contraction limit of another symmetry
\cite{3}. Our basic stable symmetry have completely
noncommutative $X_{\mu}$ and $P_{\mu}$.
We need also to see how the above perspective 
of quantum field theory. We need full dynamical pictures
which includes analysis of the corresponding algebras of
observables \cite{070}. We have an exciting long way to go.

Let us have a last word on basic quantum mechanics. 
Starting with our quantum space model, quantum mechanics
can be seen from a very different light. For instance, it begs
the question of why not sees as physical observables all
functions of the (quantum) position and momentum, as
coordinates of the quantum space/phase space. The usual
notion of  physical observables as essentially functions 
as operator $\hat{X}_i$ and $\hat{P}_i$ may be too limited.
They are only observables which have a clear classical analog.
And as such, there is no reason to restrict possible knowledge 
about the operator/observables on a state to their expectation
values either. All the other moments of the distributions are
conceptually as measurable as the  expectation values.
On a complementary point of view, the noncommutative
algebra of quantum observables (operators) can be thought
of as all continuous function $C(X)$ of a topological space $X$
is the general setting of noncommutative geometry \cite{1}.
Thinking about the quantum space as having noncommutative
coordinates, essentially $x_i\star$ and $p_i\star$ (kind of
$\hat{X}_i$ and $\hat{P}_i$)  \cite{070} is an alternative 
picture the infinite number of real number coordinates here.
We are familiar with thinking about physical quantities as
real number valued. That is no more than an assumption.
The notion of physical quantities as having values in some
other algebraic entity may be the way to go.  We are familiar
with the Bohr argument about all measurement being 
classical. That has to be the case only if we restrict results
of measurements to be real numbers. A generic 
measurement should only to a process to extract 
information from a physical system. The system being
quantum says that we may want to extract from it
quantum information rather than the familiar classical
one. We are still learning to manipulate quantum 
information well. But we can do that, we can deal with
true quantum measurements.

Dealing with physical quantities as beyond real number 
values and quantum space(time) as beyond real number
geometries is like a new Copernican revolution for
our generation to push on !

\ack
The author is partially supported by research grants 
 MOST 105-2112-M-008-017- from the MOST of Taiwan.

\smallskip

\end{document}